# Scalable and efficient separation of hydrogen isotopes using graphene-based electrochemical pumping

M. Lozada-Hidalgo[1], S. Zhang[1], S. Hu[2], A. Esfandiar[1], I. V. Grigorieva[1], A. K. Geim[1]

[1]School of Physics & Astronomy and [2]National Graphene Institute, University of Manchester

Manchester M13 9PL, UK

**Thousands of tons of isotopic mixtures are processed annually for heavy-water production and tritium decontamination. The existing technologies remain extremely energy intensive and require large capital investments. New approaches are needed to reduce the industry's footprint. Recently, micron-size crystals of graphene are shown to act as efficient sieves for hydrogen isotopes pumped through graphene electrochemically. Here we report a fully-scalable approach, using graphene obtained by chemical vapor deposition, which allows a proton-deuteron separation factor of around 8, despite cracks and imperfections. The energy consumption is projected to be orders of magnitude smaller with respect to existing technologies. A membrane based on 30 m$^2$ of graphene, a readily accessible amount, could provide a heavy-water output comparable to that of modern plants. Even higher efficiency is expected for tritium separation. With no fundamental obstacles for scaling up, the technology's simplicity, efficiency and green credentials call for consideration by the nuclear and related industries.**

**Introduction**

Separating hydrogen isotopes is a task of vast proportions. Over a thousand tons of heavy water (D$_2$O) are produced every year[1] (see Heavy water board of India at http://www.hwb.gov.in/htmldocs/general/about.asp) to supply nuclear reactors worldwide as well as for medical and research applications. The production remains expensive for two reasons. First, the low natural abundance of deuterium (0.015%) implies that huge amounts of water should be processed (the industry standard for initial enrichment is 20% of D$_2$O)[2,3]. Second, current technologies often need hundreds of stages to achieve the required degree of separation[2,4]. This means that heavy water plants are large even compared to many chemical plants[2,3]. These issues result in high capital costs and large energy consumption[2,3]. Indeed, producing 1 kg of enriched heavy water requires ~10 MWh (ref. 2), the annual energy consumption of a typical US household (U.S. Energy Information Administration). Separating tritium – hydrogen's heaviest isotope – is equally important and challenging, not least for its radioactivity[5,6]. Nuclear reactors produce tritium during their operation, which has to be continuously removed to ensure optimum performance[1,6]. Furthermore, experimental fusion facilities require tritium as a fuel[7], whereas accidents like the one at Fukushima Daiichi leave behind thousands of tons of diluted tritiated water (Ministry of Economy, Trade and Industry of Japan). The current demand stimulates search for new separation technologies that could provide higher separation factors, reduce the number of stages and minimize the energy consumption.

It has recently been shown that perfect monolayers of graphene and hexagonal boron-nitride (hBN) – impermeable to thermal atoms and molecules[8,9] – are permeable to hydrogen nuclei[10,11]. Moreover, the two-dimensional (2D) crystals could efficiently separate protons from deuterons[12]. The origin of the isotope effect was discussed in detail in ref. 12 and attributed to the energy barrier posed by the 2D crystals. The effective barrier is higher for deuterons than for protons because both are bound in their initial state to Nafion or water molecules, and zero-point oscillations at the hydrogen bond lift protons higher in energy than deuterons[12-14]. This energy difference results in different rates of thermally-activated permeation across the barrier. To investigate the isotope selectivity of the 2D materials, the experiments in ref. 12 employed small monocrystalline membranes obtained by mechanical exfoliation of bulk crystals. Industrial opportunities provided by the found isotope effect were not analyzed because this would have required membranes and fabrication methods suitable for fully scalable manufacturing, which had not been developed at the time.

The current report explores the feasibility of graphene-based electrochemical pumps for industrial-scale separation of hydrogen isotopes. We suggest that this can be achieved by roll-to-roll[15] fabrication of large-area membranes that use standard CVD graphene supported on commercially available polymer films (Nafion[16]).

**Results**

**Device Fabrication**

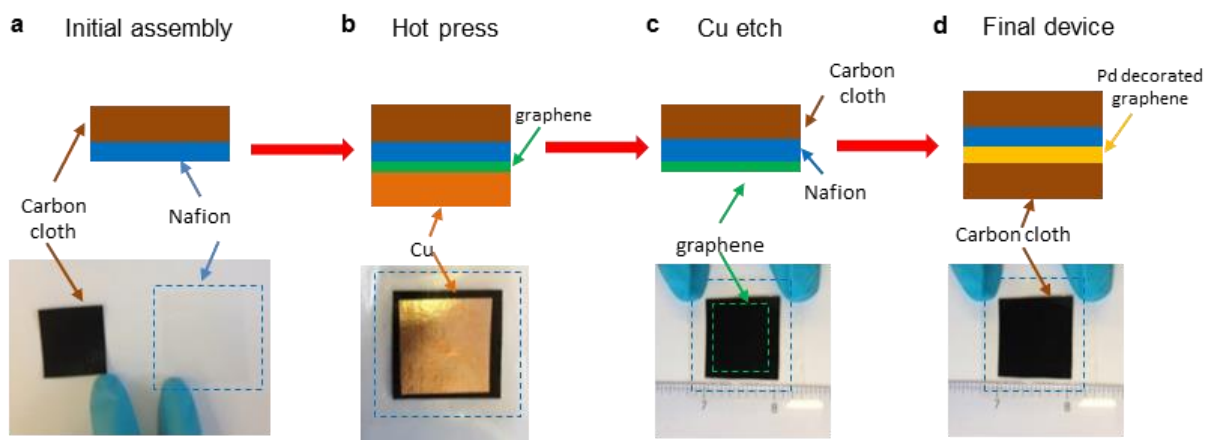

**Figure 1|Fabrication of graphene-on-Nafion membranes. a**, Optical image of a 50 μm thick Nafion film (transparent area contoured by the blue dashes) and of a piece of carbon cloth before they are stacked as shown in the sketch. **b**, Inch-size CVD graphene grown on Cu is hot pressed against Nafion. **c**, Cu foil is etched away. Graphene remains attached to Nafion (green lines indicate the graphene position). **d**, Graphene is decorated with Pd nanoparticles and sandwiched with another carbon cloth.

The graphene membranes studied in this work were fabricated as illustrated in Fig. 1. A Nafion film was first attached to a carbon cloth. Next, a Cu foil with CVD-grown graphene was hot pressed against the Nafion; the copper was subsequently etched away to release the CVD graphene onto Nafion. Then, we used electron-beam evaporation to decorate graphene with Pd nanoparticles that served as a catalyst to increase graphene's hydrogen isotope transparency[10,12]. Finally, graphene was covered with another carbon cloth, both to prevent its accidental damage and to electrically contact graphene over the entire area. To estimate the quality of graphene coverage, we examined similar membranes but without carbon-cloth electrodes (see "Device fabrication" in Methods for details). In this case, graphene becomes optically visible (Fig. 2d inset) so that cracks and holes could be analyzed under a high resolution optical microscope. This allowed us to estimate the graphene coverage as ~95%. We also checked the films using scanning electron microscopy and Raman spectroscopy (Supplementary Fig. 1), which showed that the transfer did not introduce any significant amount of microscopic defects, in agreement with optical characterization. The quality of transferred graphene films was also supported by their relatively low resistivity[15,17,18] of 780 ±160 Ω sq$^{-1}$. We limited the size of the tested membranes to 1 square inch. This was chosen for convenience only. The discussed fabrication can be extended to produce graphene-on-Nafion membranes of virtually any size, following recent advances in growth and transfer of large-area CVD graphene[15,17,18].

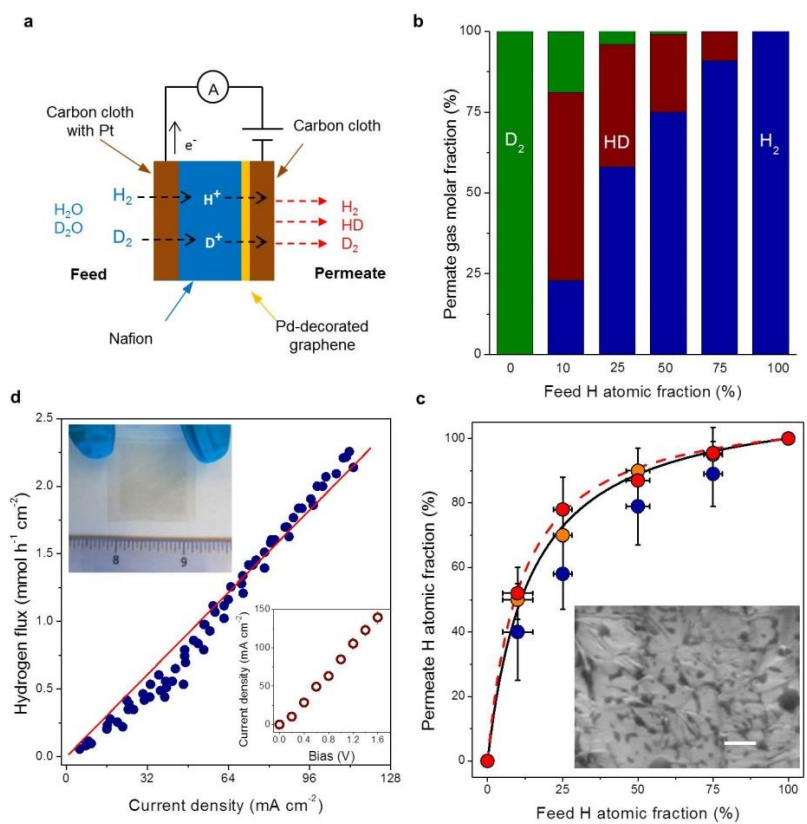

**Figure 2| Characterization of isotope separation by electrochemical pumping using CVD-graphene on Nafion**. **a**, Schematic of our setup. **b**, Gas molar fractions in the permeate, which were detected by mass spectrometry, for six different protium fractions in the feed. **c**, Protium in the permeate as a function of its concentration in the feed. Symbols: Different colors correspond to three devices. The red circles are for the measurements in (b). Solid black and dashed red curves: eq. 1 with $\alpha$ = 8 and 10, respectively. Inset: electron micrograph of CVD graphene on Nafion, scale bar 200 nm. The graphene is clearly visible

as it conforms to the rough surface of the polymer, which contains areas with different (bright and dark) electron contrast. **d**, $H_2$ flux (throughput per unit area) as a function of electric current for one of our devices with only protium in the feed. The red line is the best fit that agrees with Faraday's law within the experimental scatter. Top inset: photo of a graphene-on-Nafion membrane without carbon cloth. Graphene shows up as a darker area. Bottom inset: typical current-voltage characteristic for our devices.

**Mass spectrometry measurements**

The permeability of protons and deuterons across the graphene-on-Nafion membranes was studied using mass spectrometry[12]. To this end, the membranes were placed between two chambers, feed and permeate, and the two carbon cloths were electrically contacted as shown in Fig. 2a. In the feed, we placed a large volume of vapor-gas mixtures of $H_2O$-$H_2$ and $D_2O$-$D_2$ with a chosen atomic fraction of protium; this fraction remained fixed during a given measurement and could be changed as required. The opposite side of the membrane faced a vacuum chamber connected to a mass spectrometer. This setup represents an electrochemical pump[19], in which graphene – a mixed proton-electron conductor[10,20] – acts as both cathode and semi-permeable membrane to protons and deuterons. One of the carbon cloths is used as the anode and the other as a convenient way to electrically contact the graphene sheet (the devices also work without the latter cloth). The working mechanism of the electrochemical pump is explained in Figure 2a. By applying a voltage bias across the device, we pump protons and deuterons through Nafion and across graphene. These isotopes recombine on the permeate side to form three possible molecular species – $H_2$, $D_2$ and protium deuteride (HD) – which are detected by the mass spectrometer for each of the fixed protium fractions in the feed[12]. We simultaneously measured the electrical current through the circuit and the gas flux (throughput per unit area) for each of the gases[12].

There are two key characteristics to monitor in these measurements: isotope selectivity and hydrogen throughput. Let us discuss the selectivity first. Figure 2b shows the molar fraction of the three gases in the permeate ($H_2$, $D_2$ and HD) for different atomic fractions of protium in the feed. The data for the permeated gases can be converted directly into protium atomic fractions (see "Mass spectrometry measurements" in Methods). The latter are plotted in Figure 2c as a function of protium fraction in the feed and show that the membranes preferentially allow protons through. For example, for equal amounts of protium and deuterium in the feed (50% H atomic fraction), protium accounts for up to ≈90% of atoms in the permeate. In the previous work with 2D monocrystals[12], it was shown that the atomic fraction of protium in the feed and permeate were related by the binary-mixture equation[21]

$$H_{permeate} = \alpha H_{feed} / [1 + (\alpha-1)H_{feed}] \qquad (1)$$

with a large separation factor $\alpha \approx 10$, where $H_{feed}$ and $H_{permeate}$ are the protium atomic fractions in the feed and permeate, respectively. This relation is plotted in Fig. 2c for $\alpha \approx 10$ (dashed red curve). Fig. 2c shows that eq. (1) can also describe proton transport through our inch-sized graphene-on-Nafion membranes and, despite the presence of holes and cracks in CVD graphene, the separation factor remains high, $\alpha \approx 8$ (the best fit is shown by the solid black curve). At first glance, this may seem surprising because holes in graphene provide a much easier permeation path[10-12]. However, this confusion is resolved once one realizes that the isotopes are electrochemically pumped mostly across the areas covered with graphene, because they provide the electrical contact necessary for the pump operation. For reference, we tested devices fabricated using exactly the same procedures but without graphene, which exhibited no isotope effect, as expected (Supplementary Figure 2).

The second key characteristic of electrochemical pumping is its throughput (flux times area $A$) which for potential applications needs to be as high as possible. Fig. 2d shows the measured gas flux $\Phi$ as a function of the electric current $I$ through our devices. The dependence is accurately described by Faraday's law of electrolysis $\Phi = I/2FA$, where $F$ is Faraday's constant and the factor of 2 accounts for two atoms in each $H_2$ molecule. This means that we reached 100% Faradaic efficiency for converting electron current into hydrogen flux such that one hydrogen molecule was generated in the permeate chamber for every two electrons pushed through the electrical circuit, in agreement with previous experiments using micron-size 2D monocrystals[10,12]. This relation held even after measuring the samples for several days and in repetitive experiments.

**Energy estimations**

We typically applied biases $V \lesssim 1.5$ V to avoid electrolysis of water at the feed side. In this regime, both $I$ and $\Phi$ varied linearly with $V$ (see Fig. 2d inset). The measured flux is quite significant and, for example, at 0.5 V our electrochemical pump provided a throughput of ~0.8 mmol h$^{-1}$ cm$^{-2}$. If a membrane with an area of 30 m$^2$ is employed, this would translate into ~50,000 m$^3$ of processed gas mixtures per year (at standard temperature and pressure) or, equivalently, ~40 tons of heavy water pa. The latter number is comparable with a typical annual output of existing heavy-water production plants[1] (see also Heavy Water Board of India).

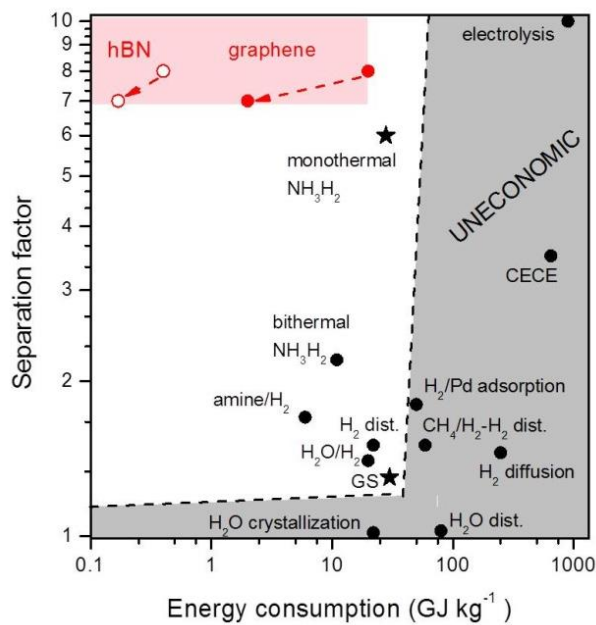

**Figure 3 | Comparison of the electrochemical pumping with other technologies available for heavy water production.** Data for the latter are adapted from ref. 2. The most common industrial methods are indicated by the black stars. Red circles: the demonstrated electrochemical pumping using CVD

graphene (solid) and an estimate for monolayer hBN on Nafion (open). The red shaded area indicates the possible parameter space if process parameters such as temperature, membrane area, driving voltage and membranes' defectiveness are changed. The arrow shows the changes expected for graphene-on-Nafion and hBN-on-Nafion with increasing temperature to 60°C. The grey area indicates the parameter space considered unsuitable for primary heavy water production. Abbreviations: "CECE" – combined electrolysis and chemical exchange, "dist." – distillation. Note that the methods that are uneconomical for primary production of heavy water are used for final enrichment.

For the known throughput and separation factor, it is straightforward to estimate the energy needed to enrich hydrogen with the natural deuterium concentration of 0.015% up to 20% required for primary heavy water production[2,4]. Despite the slight decrease in the separation factor with respect to the maximum achievable $\alpha$ ~10 for ideal graphene at room temperature[12], $\alpha$ ~8 is still high enough to minimize the amount of feed that needs to be processed. Using the Rayleigh equation[21], we find that it is required to process ~3800 moles of hydrogen feed per mole of enriched product. The total power $P$ required is given by $P = IV$. Therefore, with an operational voltage of ~0.5 V, the power density ($p=P/A$) per unit of gas flux is given by $p/\Phi = 2FV$ ~ 26 Wh mol$^{-1}$ which depends only on the applied voltage, in agreement with the data in Figure 2d. This allows for the estimate that graphene-based electrochemical pumping would involve energy costs of $E$ ~20 GJ per kg of 20% enriched heavy water.

**Comparison with existing technologies**

At this point it is instructive to compare the proposed technology with those either currently employed or considered for heavy water production. The situation is summarized in Fig. 3. It shows that the new graphene-based technology occupies a highly desirable and previously empty corner within the parameter space ($E,\alpha$). The most utilized technologies today are the Girdler-Sulfide (GS) and monothermal-$NH_3/H_2$ processes[1] (see also Heavy Water Board of India). They are characterized by $\alpha$ ~1.3 and 6, respectively, and both require $E$ ~30 GJ kg$^{-1}$. Our electrochemical pumping exhibits ($E,\alpha$) = (~20 GJ kg$^{-1}$, ~8), outperforming all the other technologies. Moreover, the above estimate for $E$ was rather conservative, so we believe that the energy costs can be reduced by at least an order of magnitude by a number of design changes. For example, the operating temperature can be increased to 60 °C, which would result in an exponential increase in graphene's proton conductivity; by a factor of 10 as proven in previous experiments[10,12]. This should lead to an equivalent reduction in $E$ (see the red arrow in Fig. 3). The separation factor would also decrease but only to ≈7 (see ref. 12), which would translate only into an increase of ~20% in the amount of feed. Another possibility is to use monolayer hBN instead of graphene. The former exhibits the same $\alpha$ as graphene but its room-temperature proton conductivity, $I/AV$, is ≈50 times higher[10,12], which translates into the same factor in reducing the energy costs. Operating hBN-based membranes at elevated temperatures[10,12] would reduce $E$ even further. Note that, unlike graphene, hBN is an electron insulator. Therefore, a nm-thick Pd film[22] would have to be evaporated on the permeate side to provide the necessary electrical conductivity but this addition should not significantly affect the performance of hBN-on-Nafion membranes, as described previously[10,12]. CVD growth of monolayer hBN has been demonstrated[23,24] but its roll-to-roll, industrial scale production has not been developed yet, probably due to the current lack of demand. An additional possibility to reduce $E$ would be to increase the membrane area while keeping the same total throughput $\Phi A$. Indeed, the above estimates were given for a fixed bias of 0.5 V. A decrease in $\Phi$ at the expense of larger $A$ would allow reduction in $V$ and proportional savings in $E$. Finally, note that the above energy estimates are based on a single stage separation process that does not exploit the advantage of optimized process designs with multiple enrichment stages[2-4]. The possible use of the processed hydrogen as an energy source for the continuous enrichment is also not taken into account, but it could cover a significant proportion of the energy costs. Therefore, our estimates should be regarded as upper-bound rather than actual energy costs. Process details, the operating temperature, the used 2D material and its area can all be adjusted to optimize an eventual design of a plant while considering its capital and operating costs (see "Energy estimations" in Supplementary Information).

**Discussion**

In conclusion, the demonstrated electrochemical pumping offers a large separation factor, extremely low energy costs, potentially small footprint, competitive throughput and rather straightforward design of industrial scale facilities. On top of this, there are no environmental issues because no toxic and corrosive substances are involved, unlike in the case of the GS and monothermal-$NH_3$/$H_2$-exchange technologies[2-4]. The new technology seems to be even more attractive for tritium decontamination and extraction. The identified mechanism behind proton-deuteron sieving implies a separation factor of ≈30 for protium-tritium separation[12]. This is in stark contrast with cryogenic distillation (the most used method) which offers a measly separation factor of ≲1.8 for hydrogen isotope mixtures[5]. Although the existing isotope separation technologies are well established, the described advantages of the graphene-based separation seem significant enough to justify rapid introduction of this disruptive technology even within the highly-conservative nuclear industry.

**References**

1. International Atomic Energy Agency. *Heavy water reactors : status and project development. International Atomic Energy Agency.* Technical reports series, no. 407 (2002).
2. Rae, H. K. *Separation of Hydrogen Isotopes, Selecting Heavy Water Processes*. (Am. Chem. Soc. 1978).
3. Miller, A. I. Heavy Water: A Manufacturers' Guide for the Hydrogen Century. *Can. Nucl. Soc. Bull.* **22,** 1–14 (2001).
4. Morris, J. W., Bebbington, W. P., Garvin, R. G. & Schroder, M. C. Heavy Water for the Savannah River Site. *Proc. Symp. Savannah River Site 50th Anniv.* (2000).
5. Vasaru, G. *Tritium Isotope Separation*. (CRC Press, 1993).
6. International Atomic Energy Agency. *Handling of Tritium-Bearing Wastes,* Technical Report Series No. 203, (1981).
7. Caporali, R. The quest for fusion energy. *J. Fusion Energy* **10,** 255–257 (1991).
8. Bunch, J. S. *et al.* Impermeable atomic membranes from graphene sheets. *Nano Lett.* **8,** 2458–62 (2008).
9. Koenig, S. P., Wang, L., Pellegrino, J. & Bunch, J. S. Selective molecular sieving through porous graphene. *Nat. Nanotechnol.* **7,** 728–32 (2012).
10. Hu, S. *et al.* Proton transport through one-atom-thick crystals. *Nature* **516,** 227–230 (2014).
11. Achtyl, J. L. *et al.* Aqueous proton transfer across single-layer graphene. *Nat. Commun.* **6,** 6539 (2015).
12. Lozada-Hidalgo, M. *et al.* Sieving hydrogen isotopes through two-dimensional crystals. *Science* **351,** 68–70 (2016).
13. Marx, D. Proton transfer 200 years after von Grotthuss: insights from ab initio simulations. *Chemphyschem* **7,** 1848–70 (2006).
14. Wiberg, K. B. The deuterium isotope effect. *Chem. Rev.* **55,** 713–743 (1955).

15. Bae, S. *et al.* Roll-to-roll production of 30-inch graphene films for transparent electrodes. *Nat. Nanotechnol.* **5,** 574–578 (2010).
16. Mauritz, K. & Moore, R. State of understanding of nafion. *Chem. Rev.* **104,** 4535–85 (2004).
17. Kim, K. S. & Al., E. Large-scale pattern growth of graphene films for stretchable transparent electrodes. *Nature* **457,** 706–710 (2009).
18. Reina, A. & Al., E. Large area, few-layer films on arbitrary substrates by chemical vapor deposition. *Nano Lett.* **9,** 30–35 (2008).
19. Iwahara, H. Hydrogen pumps using proton-conducting ceramics and their applications. *Solid State Ionics* **125,** 271–278 (1999).
20. Castro Neto, A. H., Peres, N. M. R., Novoselov, K. S. & Geim, A. K. The electronic properties of graphene. *Rev. Mod. Phys.* **81,** 109–162 (2009).
21. Gorak, A. & Sorensen, E. *Distillation: Fundamentals And Principles*. (Elsevier, 2014).
22. Nishikawa, M., Shiraishi, T. & Murakami, K. Solubility and Separation Factor of Protium-Deuterium Binary Component System in Palladium. *J. Nucl. Sci. Technol.* **33,** 504–510 (1996).
23 Kim, K. K. *et al.* Synthesis of Monolayer Boron Nitride on Cu Foil Using Chemical Vapor Deposition. *Nano Lett.* **12,** 161–166 (2012).
24. Sutter, P., Lahiri, J., Albrecht, P. & Sutter, E. Chemical Vapor Deposition and Etching of High-Quality Monolayer Hexagonal Boron Nitride Films. *ACS Nano* **9,** 7303–7309 (2011).

**Methods**

**Device fabrication**

A Nafion membrane (N212, pretreated through standard procedure) was hot pressed on a carbon cloth substrate (catalyst loading of 0.2 mg cm$^{-2}$, 20% platinum by weight on carbon support). Next, the opposite side of the Cu substrate (covered with CVD graphene on both sides) was plasma etched to remove graphene from that side. Then, the CVD graphene (still on Cu substrate) was placed on the carbon-cloth/Nafion stack and the whole assembly was hot pressed in a laminator (to prevent sticking to the hot surfaces, the two sides of the stack were protected by waxed paper and polyethylene terephthalate films). The Cu substrate was then etched in ammonium persulfate solution and the device was left in DI water to remove residues of the etchant. At this point the transfer of CVD graphene onto the Nafion substrate was complete. Finally, to facilitate proton permeation (see main text), the graphene side of the carbon-cloth/Nafion/graphene stack was decorated with Pd nanoparticles deposited by e-beam evaporation (nominally 2 nm thick) and covered by hand with another piece of carbon cloth (no catalysts).

**Mass spectrometry measurements**

Devices for mass spectrometry were placed in a custom built fuel cell set up that could be connected to our mass spectrometer. This set up was placed in between the feed and permeate chambers. In these experiments we monitored the molar fraction of the gases in the permeate chamber for a fixed deuterium atomic fraction in the feed. The feed chamber was then evacuated and the vapor-gas mixtures of $H_2O$-$H_2$ and $D_2O$-$D_2$ of known molar fraction were allowed to flow into the feed. Because of the large volume of the feed chamber compared to what our devices can process throughout the whole measurement, the chamber effectively represents an infinite reservoir of gases at a fixed deuterium atomic fraction. We normally used 10% $H_2$-$D_2$ in Ar at 100% $H_2O$-$D_2O$ humidity. The only exception to this was when we specifically studied the throughput across the membranes (Fig. 2 in the main text), in which case we used 100% $H_2$ at 100% $H_2O$ humidity. To ensure that the protium-deuterium fraction in Nafion was the same as in the chamber itself, the gas mixture in the feed chamber was fixed and the device was biased at ~1V for 20 min to allow for the protium-deuterium mixture in Nafion to equilibrate with the one in the chamber. This method has been used before in other studies and we further verified its effectiveness with our reference devices without graphene (Supplementary Figure 2). To calibrate the three gas signals from the mass spectrometer, we first used a known leak to calibrate the $H_2$ and $D_2$ flux under 100% protium and 100% deuterium atomic fraction conditions, respectively. We note then that at *RT* the molar fraction of HD ([HD]) is related to the $H_2$ and $D_2$ molar fractions ([$H_2$] and [$D_2$], respectively) in the permeate chamber via the law of mass action: $[HD]^2/\{[H_2][D_2]\}=K(T)$ where *K*(*T*) is a constant. Therefore, using a control sample (no graphene), we calibrated the HD response using this law.